\documentclass[
 reprint,
 superscriptaddress,
 amsmath,amssymb,
 aps,
 prl
]{revtex4-2}

\usepackage{graphicx}
\usepackage{dcolumn}
\usepackage{bm}

\usepackage{hyperref} 

\begin{document}

\title{
Subthermal Mean Transverse Energies Induced by Electron Refraction on the Jump in Mass at the Surface of Multialkali Photocathodes
}

\author{S. A. Rozhkov}
\email{rozhkovs@isp.nsc.ru}
\affiliation{Rzhanov Institute of Semiconductor Physics, Siberian Branch,
Russian Academy of Sciences, Novosibirsk 630090, Russia}
\affiliation{Novosibirsk State University, Novosibirsk 630090 Russia}

\author{V. V. Bakin}
\affiliation{Rzhanov Institute of Semiconductor Physics, Siberian Branch,
Russian Academy of Sciences, Novosibirsk 630090, Russia}

\author{H. E. Scheibler}
\affiliation{Rzhanov Institute of Semiconductor Physics, Siberian Branch,
Russian Academy of Sciences, Novosibirsk 630090, Russia}
\affiliation{Novosibirsk State University, Novosibirsk 630090 Russia}
\affiliation{Novosibirsk State Technical University, Novosibirsk 630073 Russia}

\author{V. S. Rusetsky}
\affiliation{Rzhanov Institute of Semiconductor Physics, Siberian Branch,
Russian Academy of Sciences, Novosibirsk 630090, Russia}
\affiliation{JSC ``EKRAN FEP", Novosibirsk 630060, Russia}

\author{D. V. Gorshkov}
\affiliation{Rzhanov Institute of Semiconductor Physics, Siberian Branch,
Russian Academy of Sciences, Novosibirsk 630090, Russia}
\affiliation{Novosibirsk State University, Novosibirsk 630090 Russia}

\author{D. A. Kustov}
\affiliation{Rzhanov Institute of Semiconductor Physics, Siberian Branch,
Russian Academy of Sciences, Novosibirsk 630090, Russia}

\author{V. A. Golyashov}
\affiliation{Rzhanov Institute of Semiconductor Physics, Siberian Branch,
Russian Academy of Sciences, Novosibirsk 630090, Russia}
\affiliation{Novosibirsk State University, Novosibirsk 630090 Russia}
\affiliation{Synchrotron Radiation Facility SKIF, Boreskov Institute of
Catalysis, Siberian Branch, Russian Academy of Sciences, Koltsovo 630559, Russia}

\author{V. L. Alperovich}
\affiliation{Rzhanov Institute of Semiconductor Physics, Siberian Branch,
Russian Academy of Sciences, Novosibirsk 630090, Russia}
\affiliation{Novosibirsk State University, Novosibirsk 630090 Russia}

\author{O. E. Tereshchenko}
\email{teresh@isp.nsc.ru}
\affiliation{Rzhanov Institute of Semiconductor Physics, Siberian Branch,
Russian Academy of Sciences, Novosibirsk 630090, Russia}
\affiliation{Novosibirsk State University, Novosibirsk 630090 Russia}
\affiliation{Synchrotron Radiation Facility SKIF, Boreskov Institute of
Catalysis, Siberian Branch, Russian Academy of Sciences, Koltsovo 630559, Russia}

\date{\today}

\begin{abstract}
The search for photocathode materials with low mean transverse energies (MTEs) and, hence, low intrinsic emittance is of crucial importance for various fields of particle and solid state physics. Here, we demonstrate that polycrystalline multialkali Na$_{2}$KSb(Cs,Sb) photocathodes with negative effective electron affinity (NEA) have MTE values at room temperature by a factor of 2 lower than those of monocrystalline \textit{p}-GaAs(Cs,O) photocathodes. These low MTE values are due to the electron refraction on the jump in mass, between a small effective mass in Na$_{2}$KSb and free electron mass in vacuum. It is proved that, at the NEA state, up to half of photoelectrons are emitted in a narrow-angle cone with the fractional MTE of 9\,meV at room temperature. We also showed that the transition from NEA to positive effective affinity results in the subthermal total MTE of the Na$_{2}$KSb(Cs,Sb) photocathode, along with quantum efficiency of about 10$^{-2}$. The physical reasons for the manifestation of the refraction effect in multialkali photocathodes are discussed, opening up opportunities for the development of high-brightness and ultracold robust electron sources.
\end{abstract}
                              
\maketitle

The discovery of photoemission and its explanation by Einstein more than a century ago was a major stimulus for the creation of the quantum picture of the world, for the development of the most powerful method to study the electronic structure of condensed matter---angle-resolved photoemission spectroscopy (ARPES) \cite{Sobota2021}, as well as for the development of a number of photoemission devices: photomultipliers, image intensifiers, solid-state sources of cold and spin-polarized electrons \cite{Bell1973, Dunham2013, Orlov2004, Mamaev2008}. Despite decades of research and application developments, the parameters of the devices are still far from theoretical limits. Photocathodes for the generation of high-brightness and ultracold electron beams are needed in various fields of science, such as free electron lasers \cite{DiMitri2014}, cryogenic ion storage rings \cite{Novotny2019}, ultrafast electron diffraction \cite{Filippetto2022} and energy-loss spectroscopy \cite{Pomarico2018}.

\begin{figure}
\includegraphics[width=1.0\linewidth]{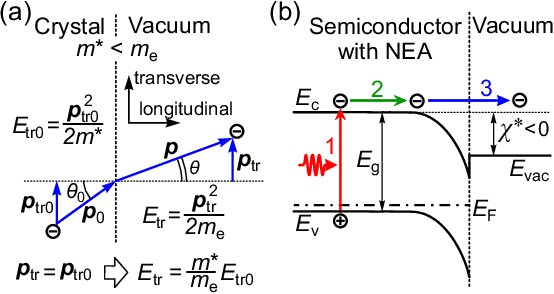}
\caption{\label{Fig.1}
(a) Electron refraction on the jump in mass at the ideal crystalline solid-vacuum interface and the reduction of the electron transverse energy $E_{\text{tr}}$ in vacuum due to the conservation of transverse component of electron momentum $\textit{\textbf{p}}_{\text{tr}} = \textit{\textbf{p}}_{\text{tr0}}$. (b) Energy-band diagram of a semiconductor photocathode with NEA ($\chi^{*} < 0$) and a three-step photoemission process: (1) photoexcitation, (2) transport to the emitting surface and (3) escape into vacuum. $E_{\text{c}}$ and $E_{\text{v}}$ are the conduction band bottom and the valence band top in the active layer, respectively, $E_{\text{g}}$ is the band gap, $E_{\text{F}}$ is the Fermi level, $E_{\text{vac}}$ is the vacuum level.}
\end{figure}

Along with quantum efficiency (QE), a key factor that limits the brightness of an electron beam is the mean transverse energy (MTE) of electrons emitted by a photocathode \cite{Musumeci2018}. Despite considerable efforts in the search for photocathodes with low MTE, up to now, the experimental MTE values reported in the literature exceeded the thermal energy $k_{B}T$ or were close to $k_{B}T$; here $T$ is photocathode temperature, $k_{B}$ is Boltzmann constant \cite{Rodway1986, Orlov2004, Karkare2014, Feng2015, Karkare2020, Musumeci2018, Nangoi2021, Schroeder2025}. However, theoretically, it is possible to overcome this ``thermal limit” and obtain subthermal $\text{MTE} < k_{B}T$ due to the conservation of transverse component of a momentum upon the electron transfer through the interface between a crystalline solid and vacuum [Fig.~\ref{Fig.1}(a)]. In this case, similar to the Snell’s law in optics, the refraction of trajectories occurs on the jump in mass, between small effective mass $m^{*}$ in a crystal and free electron mass $m_{\text{e}}$ in vacuum, which should result in a narrow cone emission with a reduced transverse energy of emitted electrons $E_{\text{tr}} = (m^{*}/m_{\text{e}}) E_{\text{tr0}}$, where $E_{\text{tr0}}$ is the transverse energy inside a crystal with a parabolic conduction band \cite{Bell1973, Vergara1996, Rickman2013, *Rickman2014, Alperovich2021}. For the isotropic thermal distribution of photoelectrons in a conduction band, the transverse energy distribution of emitted electrons $N_{\text{e}}(E_{\text{tr}})$ should be as follows:
\begin{equation} \label{Eq.1}
N_{\text{e}}(E_{\text{tr}}) \propto \text{exp} \{ -E_{\text{tr}} / [ k_{B}T (m^{*} / m_{\text{e}}) ] \},
\end{equation}
with MTE of emitted electrons as low as $k_{B}T (m^{*}/m_{\text{e}})$ \cite{Bell1973}.

The electron refraction on the jump in mass was most extensively studied in \textit{p}-GaAs(Cs,O) and other III-V semiconductor photocathodes with negative effective electron affinity (NEA) due to both small electron effective mass ($m^{*}_{\text{GaAs}} = 0.067 m_{\text{e}} $) and practical importance for low-light imaging \cite{Bell1973} and generation of spin-polarized electrons \cite{Mamaev2008}. Narrow angular distributions of electrons emitted from GaAs(Cs,O) photocathodes were reported in Refs.~\cite{Pollard1974, Lee2007} in agreement with the electron refraction on the jump in mass. However, the techniques for measuring angular distributions of low energy ($< 1$\,eV) electrons, which were used in Refs.~\cite{Pollard1974, Lee2007}, were later criticized \cite{Bradley1977, Lee2015}, and electron emission into a broad solid angle with MTE values in the range of 25--120\,meV was observed in most papers on GaAs(Cs,O) NEA photocathodes \cite{Rodway1986, Vergara1996, Orlov2001, Bakin2003, Orlov2004, Karkare2011, Karkare2014, Jones2021}. Bakin \textit{et al.} \cite{Bakin2003} proved experimentally that refraction occurred only for a small ($\sim 5\%$) group of photoelectrons, which emitted from GaAs(Cs,O) photocathode ballistically (without diffuse scattering at the surface) with the ``fractional" MTE of about 1\,meV at $T = 77$\,K, while the total MTE was about 100\,meV. The domination of emission in a broad solid angle was attributed to the surface roughness \cite{Bradley1977, Karkare2011}, electron diffuse scattering in the amorphous (Cs,O) activation layer \cite{Rodway1986, Vergara1996, Karkare2017} and electron-phonon interaction \cite{Schroeder2025}. Blurring of low-energy electron diffraction (LEED) patterns under Cs and $\text{O}_{2}$ deposition on GaAs surfaces confirms strong diffuse scattering in the (Cs,O) activation layer \cite{Goldstein1975, Bakin2007}. Berger \textit{et al.} \cite{Berger2012} studied the correlation between the transverse momentum (and, hence, MTE) and effective mass $m^{*}$ due to the refraction of emitted electrons on clean GaSb and InSb surfaces by the excited-state thermionic emission, although the obtained MTE exceeded the thermal energy $k_{B}T$.

The dependence of the transverse momentum of emitted electrons on effective mass $m^{*}$ was considered for robust metal photocathodes \cite{Rickman2013, *Rickman2014, Li2015}, and, again, the MTE thermal limit was not overcome neither in polycrystalline, nor in monocrystalline metals \cite{Feng2015, Karkare2017, Karkare2020}. MTE close to, but exceeding $k_{B}T$ was obtained by decreasing photon energy $\hbar\omega$ down to work function $W$. At these photon energies the QE values are in the range of $\sim 10^{-5}$--$10^{-8}$ \cite{Dowell2009, Karkare2017, Karkare2020}; this restricts the use of metals as electron sources in many applications.

Photocathodes based on the alkali antimonide semiconductor compounds, like Cs$_{3}$Sb, K$_{2}$CsSb, Na$_{2}$KSb and Na$_{2}$KSb(Cs,Sb), are characterized by an optimal combination of relatively high QE values (compared to metals), along with good robustness (compared to fragile GaAs(Cs,O) photocathodes) \cite{Spicer1958, Dunham2013, Petrushina2020}. Therefore, the opportunities to obtain alkali antimonide photocathodes with low MTE values were actively studied in recent decades. Similarly to metals, for alkali antimonides, MTE values close to thermal energy were reached below the photoemission threshold \cite{Cultrera2016, Musumeci2018, Dube2025}.

\begin{figure*}
\includegraphics[width=1.0\linewidth]{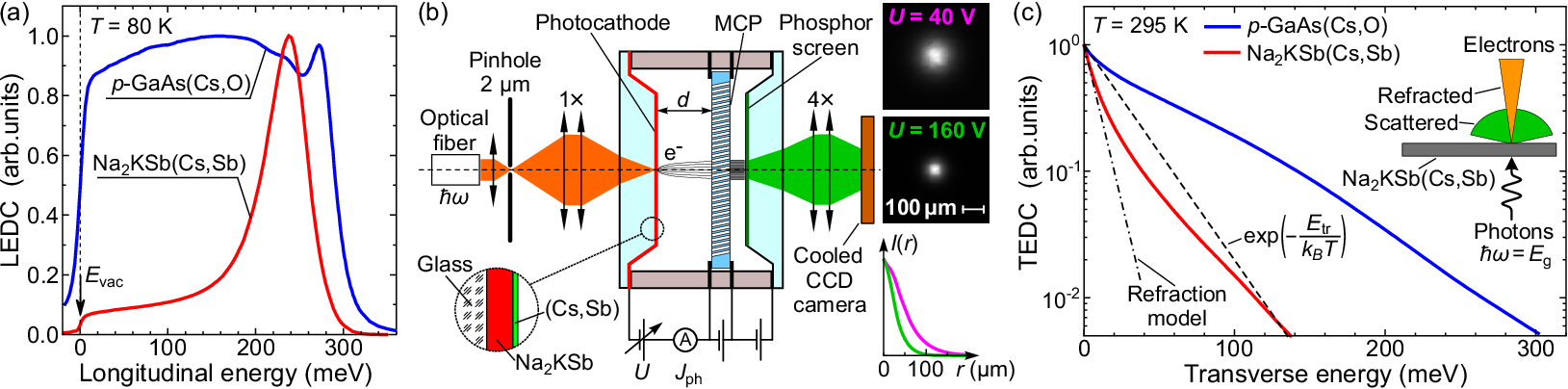}
\caption{\label{Fig.2}
(a) Longitudinal energy distribution curves of Na$_{2}$KSb(Cs,Sb) and \textit{p}-GaAs(Cs,O) photocathodes measured at $T = 80$\,K and $\hbar\omega = E_{\text{g}}$. Vacuum level $E_{\text{vac}}$ is marked with an arrow. (b) A schematic representation of a vacuum tube and apparatus for the measurements of both longitudinal and transverse electron distributions. $d$ is the photocathode-MCP gap, $U$ is the photocathode-MCP voltage, $J_{\text{ph}}$ is the photoemission current. The measured images of the electron beam emitted by the Na$_{2}$KSb(Cs,Sb) photocathode and the corresponding radial intensity distributions $I(r)$ are shown for two accelerating voltages. (c) Transverse energy distribution curves of Na$_{2}$KSb(Cs,Sb) and \textit{p}-GaAs(Cs,O) photocathodes measured at $T = 295$\,K and $\hbar\omega = E_{\text{g}}$. The thermal transverse distribution and refraction model distribution [Eq.(\ref{Eq.1})] for $m^{*} = 0.35 m_{\text{e}}$ are shown by dashed and dash-dotted lines, respectively. Inset: Schematic angular distributions of electrons emitted partly into a narrow cone due to the refraction on the jump in mass and partly into a wide solid angle due to diffuse scattering.}
\end{figure*}

Recently, Rusetsky \textit{et al.} \cite{Rusetsky2022} have discovered the optical orientation and emission of spin-polarized electrons from the multialkali Na$_{2}$KSb(Cs,Sb) NEA photocathodes, which consist of $\sim 100$\,nm thick polycrystalline Na$_{2}$KSb active layers and thin ($\sim 1$\,nm) (Cs,Sb) activation layers \cite{Spicer1958, Ninomiya1969, Dolizy1988, Erjavec1997}. These results are promising for the development of robust spin-polarized electron sources. Our following studies of multialkali photocathodes proved that the energy band diagrams and emission mechanisms of \textit{p}-GaAs(Cs,O) and  Na$_{2}$KSb(Cs,Sb) NEA photocathodes are similar [see Fig.~\ref{Fig.1}(b)] \cite{Rusetsky2022, Rozhkov2024, Tereshchenko2025}. However, along with the similarities, it turned out that the shapes of the longitudinal energy distribution curves (LEDCs) of Na$_{2}$KSb(Cs,Sb) and GaAs(Cs,O) photocathodes are qualitatively different at cryogenic temperatures [Fig.~\ref{Fig.2}(a)]. This difference suggests that multialkali photocathodes emit electrons predominantly along the normal to the surface, and that diffuse scattering at the Na$_{2}$KSb(Cs,Sb)-vacuum interface is weaker than at the GaAs(Cs,O)-vacuum interface \cite{Rozhkov2024}. Therefore, the multialkali photocathodes seem promising for the observation of the electron refraction on the jump in mass and obtaining electron beams with subthermal MTEs.

In this Letter, we prove this hypothesis by the measurements of the transverse energy distributions of electrons emitted from the multialkali and \textit{p}-GaAs(Cs,O) photocathodes. We found that about half of all photoelectrons emit from Na$_{2}$KSb(Cs,Sb) ballistically and undergo refraction on the jump in mass. Due to the electron refraction, the MTE values of Na$_{2}$KSb(Cs,Sb) photocathodes are by a factor of 2 smaller than those of GaAs(Cs,O) photocathodes. For multialkali photocathodes with positive effective affinity, a record low subthermal $\text{MTE} = 18$\,meV at room temperature was reached. The obtained results prove the potential of the Na$_{2}$KSb(Cs,Sb) photocathodes for developing robust sources of spin-polarized electrons with both high QE and low intrinsic emittance.

The experiments were performed on semitransparent Na$_{2}$KSb(Cs,Sb) \cite{Rusetsky2022, Rozhkov2024} and \textit{p}-GaAs(Cs,O) \cite{Bakin2007, Tereshchenko2017, Golyashov2020} NEA photocathodes, which were sealed in vacuum tubes along with microchannel plates (MCP) and phosphor screens [Fig.~\ref{Fig.2}(b)]. The tubes ensured the photocathodes stability and allowed us to measure both LEDCs and transverse energy distribution curves (TEDCs). To measure the LEDCs, retarding voltages $U$ were applied between the photocathode and MCP. LEDCs were obtained by measuring the derivatives of photocurrent-voltage characteristics $J_{\text{ph}}(U)$ \cite{Rozhkov2024}.

The TEDC measurements are based on the transverse expansion of an electron beam in a uniform accelerating electric field \cite{Feng2015, Jones2022, Dube2025}. At sufficiently high accelerating voltages $U$, the transverse electron displacement $r$ during the motion across the photocathode-MCP vacuum gap (with thickness $d$ of 1.6 and 3.2\,mm) is proportional to the square root of transverse energy $E_{\text{tr}}$. The photocathode was illuminated by a small light spot of $\sim 5~\mu\text{m}$ in diameter. The radial distribution of the photoemission current $I_{\text{e}}(r)$ was amplified by the MCP (with pores of $6~\mu\text{m}$ in diameter) and transferred to the phosphor screen. The light intensity distribution on the screen $I(r)$, which is proportional to $I_{\text{e}}(r)$, was measured at various $U$ by a cooled charge-coupled device (CCD) camera, as illustrated in Fig.~\ref{Fig.2}(b). The electron transverse energy $E_{\text{tr}}$ was determined with a relative error of 10\%. The details of the measurement procedure and data processing are described in Ref.~\cite{Jones2022}.

Transverse energy distributions for Na$_{2}$KSb(Cs,Sb) and GaAs(Cs,O) photocathodes with close values of NEA ($\chi^{*} = -100$ and $-110$\,meV, respectively) measured at $T = 295$\,K for near-bandgap excitation are shown in Fig.~\ref{Fig.2}(c). It is seen that the GaAs(Cs,O) distribution is much broader ($\text{MTE} = 63$\,meV) than that of Na$_{2}$KSb(Cs,Sb) ($\text{MTE} = 24$\,meV). The broad transverse distribution of electrons from GaAs(Cs,O), with the MTE value of about half of the NEA modulus $|\chi^{*}|$, agrees with previous observations \cite{Orlov2001, Jones2021} and confirms strong diffuse scattering at the GaAs(Cs,O)-vacuum interface that masks the electron refraction on the jump in mass. 

Despite practically the same NEA values, the transverse energy distribution of the Na$_{2}$KSb(Cs,Sb) photocathode is much narrower, with the MTE of 24\,meV, i.e., slightly below the thermal limit. This indicates a much weaker electron scattering at the Na$_{2}$KSb(Cs,Sb)-vacuum interface. Also, it is seen that the Na$_{2}$KSb(Cs,Sb) TEDC cannot be approximated by a single exponent. Therefore, in this case, the characterization of the transverse distribution by a single MTE value does not adequately reflects the physics of photoemission. The measured TEDC can be approximated by the sum of two exponential contributions corresponding to significantly different values of the fractional MTEs. The steep, low-energy contribution contains about 40\% of all emitted electrons and corresponds to the fractional MTE of approximately $E_{1} = 9$\,meV, which is by a factor of 3 smaller than the “thermal limit”. This fact directly proves the electron refraction on the jump in mass for a considerable part of electrons emitted by the Na$_{2}$KSb(Cs,Sb) photocathode. According to the simple model of refraction [Eq.(\ref{Eq.1})], this value of $E_{1}$ corresponds to electron effective mass $m^{*} = 0.35 m_{\text{e}}$. So far, electron effective mass in Na$_{2}$KSb has not been determined experimentally. The theoretical value of $m^{*} = 0.36 m_{\text{e}}$ was reported in Ref.~\cite{Sharma2019}. Our approximation of the Na$_{2}$KSb conduction band dispersion calculated in Ref.~\cite{Rusetsky2022} yields a smaller value $m^{*} = 0.15 m_{\text{e}}$. Thus, the effective mass determined from the measured TEDC is within the range of available calculated values. It should be noted that the low-energy part of TEDC can be additionally broadened by possible deviations in the Na$_{2}$KSb crystallite orientations, similar to the TEDC broadening upon emission from a rough photocathode \cite{Bradley1977, Karkare2011}. Therefore, the experimentally obtained value $m^{*} = 0.35 m_{\text{e}}$ should be considered as the upper limit for the electron effective mass in Na$_{2}$KSb. The second, less steep, high-energy contribution to the TEDC of Na$_{2}$KSb(Cs,Sb) with the fractional MTE $E_{2} = 34$\,meV can be attributed to electrons which undergo momentum scattering at the crystallite boundaries and (Cs,Sb) layer imperfections. 

It should be noted that the TEDC of the GaAs(Cs,O) photocathode also contain a small but distinct beak-like feature corresponding to a small fractional MTE at low transverse energies [Fig.~\ref{Fig.2}(c)]. This feature can be attributed to the group of ballistic electrons which undergo refraction \cite{Bakin2003}. However, unlike multialkali photocathodes, in GaAs(Cs,O) this group contains only $\sim 3\%$ of all electrons and, therefore, does not lead to a significant reduction in the total MTE. 

Thus, for near-bandgap excitation, TEDCs of both GaAs(Cs,O) and Na$_{2}$KSb(Cs,Sb) NEA photocathodes contain contributions from two groups of electrons [see inset in Fig.~\ref{Fig.2}(c)]: (i) ballistic electrons, which undergo refraction on the jump in mass at the photocathode-vacuum interface and yield small subthermal fractional MTE; and (ii) diffusely scattered electrons that yield large fractional $\text{MTE} \sim |\chi^{*}|/2$. In GaAs(Cs,O), due to diffuse scattering in the amorphous (Cs,O) activation layer, the second group (ii) dominates, while the first group (i) makes up only a few percent of the total number of emitted electrons; so, it is hardly possible to obtain a low subthermal MTE value of the total distribution. On the contrary, in Na$_{2}$KSb(Cs,Sb) NEA photocathodes, both groups yield comparable contributions to the transverse distributions; therefore, it is feasible to obtain much smaller MTE values than in GaAs(Cs,O). The MTE of Na$_{2}$KSb(Cs,Sb) may be further decreased by setting $|\chi^{*}|$ to a near-zero value and, thus, reducing the contribution (ii) of scattered electrons.

\begin{figure}
\includegraphics[width=1.0\linewidth]{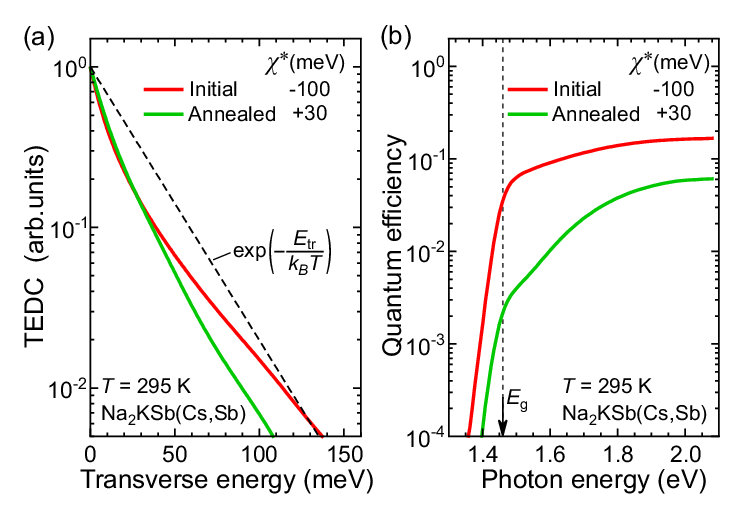}
\caption{\label{Fig.3}
(a) Transverse energy distribution curves of the Na$_{2}$KSb(Cs,Sb) photocathode measured at $T = 295$\,K and $\hbar\omega = E_{\text{g}}$ before and after annealing. The thermal transverse distribution is shown by the dashed line. (b) Photoemission quantum efficiency spectra of the Na$_{2}$KSb(Cs,Sb) photocathode measured at $T = 295$\,K before and after annealing. Band gap $E_{\text{g}}$ is marked with an arrow.}
\end{figure}

To test this opportunity, we annealed a vacuum tube with a multialkali photocathode at moderate temperature $T = 390$\,K for 2\,h in order to increase the work function due to a partial desorption of the activation layer. The analysis of the measured LEDCs showed that, as a result of the annealing, the effective affinity changed from negative $\chi^{*} = -100$\,meV to positive $\chi^{*} = +30$\,meV \cite{Rozhkov2024}. The TEDCs and QE spectra of the Na$_{2}$KSb(Cs,Sb) photocathode before and after annealing are shown in Figs.~\ref{Fig.3}(a) and \ref{Fig.3}(b), respectively. As expected, annealing led to a reduction in the high-energy contribution of scattered electrons in TEDC. This reduction allowed us to decrease the total MTE from 24\,meV to a record low subthermal value of 18\,meV at room temperature. It is seen that relatively high QE of about $10^{-2}$ can be reached along with the subthermal MTE.

It should be noted that, as a result of annealing, the width of the low-energy contribution component increased by $\sim 15\%$ probably due to the surface roughening during a partial desorption of the activation layer. Thus, annealing is not an optimal technique to reduce MTE. The partial activation with Cs and Sb aimed at near-zero effective electron affinity seems to be a better way to further reduce MTE values of multialkali photocathodes.

The measurements of the transverse EDCs prove that multialkali photocathodes with polycrystalline active layers have superior directional photoemission properties, compared to other photocathodes, including those based on GaAs(Cs,O) with structurally perfect monocrystalline active layers. We believe that this surprising and advantageous fact is based on a number of non-obvious properties of Na$_{2}$KSb(Cs,Sb) photocathodes. First, the crystallites of Na$_{2}$KSb active layers should be preferentially oriented in the photocathode plane. The available data on the specific orientation are controversial: the predominant crystallographic plane, parallel to the photocathode surface, was reported to be (210) \cite{Ninomiya1969} and (111) \cite{Dolizy1988}. Second, the emitting surfaces of Na$_{2}$KSb crystallites should be atomically smooth. Although the smoothness of Na$_{2}$KSb and Na$_{2}$KSb(Cs,Sb) surfaces was not yet proved by direct atomic-probe techniques, the smooth surfaces of alkali antimonides with sub-nm roughness were successfully obtained for Cs-Sb films \cite{Feng2017, Parzyck2023}. Third, the (Cs,Sb) activation layer formed on top of the Na$_{2}$KSb active layer must be structurally ordered since the conservation of transverse momentum and, consequently, the electron refraction takes place only if the lateral periodicity of the crystal is ensured on the scale of the de Broglie wavelength of emitting electrons. One can assume that an ordered activation layer is formed due to the isomorphism between the Na$_{2}$KSb active layer and the (Cs,Sb) activation layer \cite{Erjavec1997}, in contrast to the amorphous (Cs,O) layer on the GaAs surface. Unlike (Cs,O) activation layers, which are deposited at room temperature, (Cs,Sb) layers are deposited at elevated temperatures of about 420\,K \cite{Erjavec1997, Ninomiya1969}, and this also contributes to the formation of ordered activation layers. Again, the polycrystalline structure of Na$_{2}$KSb layers can significantly relax the lattice-matching condition for the growth of ordered (Cs,Sb) layers on each Na$_{2}$KSb crystallite, because the critical thickness of the pseudomorphic epitaxial layer on a lattice-mismatched substrate increases with reducing the lateral size of the substrate, which, in our case, is equal to a typical crystallite size of $\sim 100$\,nm \cite{Luryi1986, Ninomiya1969}.

A promising way for further improvement of the directional photoemission properties of the photocathodes consists in the growth of monocrystalline Na$_{2}$KSb films on suitable substrates. The feasibility of this approach for alkali antimonide photocathodes has been recently demonstrated by the successful growth of monocrystalline Cs-Sb films \cite{Parzyck2022}.

In conclusion, the electron refraction on the jump in mass at the solid-vacuum interface is experimentally observed in multialkali Na$_{2}$KSb(Cs,Sb) NEA photocathodes by measuring the transverse energy distributions of emitted electrons. The refraction effect allowed us to obtain record low subthermal values of MTE at room temperature along with QE of about $10^{-2}$. It was found that, in the multialkali photocathodes, at near-bandgap excitation, about a half of photoelectrons undergo the refraction on the jump in mass and emit ballistically with the fractional MTE of approximately 9\,meV, while the other half undergoes diffuse scattering upon emission with the fractional MTE proportional to the NEA modulus; averaging over these two contributions results in the total MTE of 24\,meV. Annealing the Na$_{2}$KSb(Cs,Sb) photocathode at 390\,K led to the transition from NEA to positive affinity and, thus, reduced the total MTE to the subthermal value of 18\,meV at room temperature. The comparison of transverse distributions of monocrystalline GaAs(Cs,O) and polycrystalline Na$_{2}$KSb(Cs,Sb) NEA photocathodes revealed that, in Na$_{2}$KSb(Cs,Sb) photocathodes, the ballistic contribution is by an order of magnitude higher and the MTE is by a factor of 2 lower than in the GaAs(Cs,O) photocathodes with close NEA values. This sharp difference is due to much weaker electron diffuse scattering in the (Cs,Sb) activation layers, compared to amorphous (Cs,O) activation layers, which indicates an ordered structure of the (Cs,Sb) layer. Among possible reasons for the formation of ordered (Cs,Sb) layers are the isomorphism between the active Na$_{2}$KSb and activation (Cs,Sb) layers, elevated temperature of the formation and the pseudomorphic growth of (Cs,Sb) layers on the Na$_{2}$KSb crystallites. The \textit{in situ} studies of the atomic and electronic structure of Na$_{2}$KSb(Cs,Sb) photocathodes are currently under way in our lab. Our preliminary LEED and ARPES results are consistent with the findings presented in this Letter and will hopefully facilitate the optimization of growth and activation processes to further reduce the MTE of multialkali Na$_{2}$KSb(Cs,Sb) photocathodes.

\begin{acknowledgments}

The authors acknowledge the support from the Russian Science Foundation (Grant No. 25-62-00004), SRF SKIF Boreskov Institute of Catalysis (FWUR-2024-0040) and ISP SB RAS. The authors would like to thank Professor A. S. Terekhov for stimulating discussions.

\end{acknowledgments}

\providecommand{\noopsort}[1]{}\providecommand{\singleletter}[1]{#1}%


\begin{thebibliography}{52}%
\makeatletter
\providecommand \@ifxundefined [1]{%
 \@ifx{#1\undefined}
}%
\providecommand \@ifnum [1]{%
 \ifnum #1\expandafter \@firstoftwo
 \else \expandafter \@secondoftwo
 \fi
}%
\providecommand \@ifx [1]{%
 \ifx #1\expandafter \@firstoftwo
 \else \expandafter \@secondoftwo
 \fi
}%
\providecommand \natexlab [1]{#1}%
\providecommand \enquote  [1]{``#1''}%
\providecommand \bibnamefont  [1]{#1}%
\providecommand \bibfnamefont [1]{#1}%
\providecommand \citenamefont [1]{#1}%
\providecommand \href@noop [0]{\@secondoftwo}%
\providecommand \href [0]{\begingroup \@sanitize@url \@href}%
\providecommand \@href[1]{\@@startlink{#1}\@@href}%
\providecommand \@@href[1]{\endgroup#1\@@endlink}%
\providecommand \@sanitize@url [0]{\catcode `\\12\catcode `\$12\catcode
  `\&12\catcode `\#12\catcode `\^12\catcode `\_12\catcode `\%12\relax}%
\providecommand \@@startlink[1]{}%
\providecommand \@@endlink[0]{}%
\providecommand \url  [0]{\begingroup\@sanitize@url \@url }%
\providecommand \@url [1]{\endgroup\@href {#1}{\urlprefix }}%
\providecommand \urlprefix  [0]{URL }%
\providecommand \Eprint [0]{\href }%
\providecommand \doibase [0]{https://doi.org/}%
\providecommand \selectlanguage [0]{\@gobble}%
\providecommand \bibinfo  [0]{\@secondoftwo}%
\providecommand \bibfield  [0]{\@secondoftwo}%
\providecommand \translation [1]{[#1]}%
\providecommand \BibitemOpen [0]{}%
\providecommand \bibitemStop [0]{}%
\providecommand \bibitemNoStop [0]{.\EOS\space}%
\providecommand \EOS [0]{\spacefactor3000\relax}%
\providecommand \BibitemShut  [1]{\csname bibitem#1\endcsname}%
\let\auto@bib@innerbib\@empty
\bibitem [{\citenamefont {Sobota}\ \emph {et~al.}(2021)\citenamefont {Sobota},
  \citenamefont {He},\ and\ \citenamefont {Shen}}]{Sobota2021}%
  \BibitemOpen
  \bibfield  {author} {\bibinfo {author} {\bibfnamefont {J.~A.}\ \bibnamefont
  {Sobota}}, \bibinfo {author} {\bibfnamefont {Y.}~\bibnamefont {He}},\ and\
  \bibinfo {author} {\bibfnamefont {Z.-X.}\ \bibnamefont {Shen}},\ }\href
  {https://doi.org/10.1103/RevModPhys.93.025006} {\bibfield  {journal}
  {\bibinfo  {journal} {Rev. Mod. Phys.}\ }\textbf {\bibinfo {volume} {93}},\
  \bibinfo {pages} {025006} (\bibinfo {year} {2021})}\BibitemShut {NoStop}%
\bibitem [{\citenamefont {Bell}(1973)}]{Bell1973}%
  \BibitemOpen
  \bibfield  {author} {\bibinfo {author} {\bibfnamefont {R.~L.}\ \bibnamefont
  {Bell}},\ }\href@noop {} {\emph {\bibinfo {title} {Negative Electron Affinity
  Devices}}}\ (\bibinfo  {publisher} {Clarendon Press, Oxford},\ \bibinfo
  {year} {1973})\BibitemShut {NoStop}%
\bibitem [{\citenamefont {Dunham}\ \emph {et~al.}(2013)\citenamefont {Dunham},
  \citenamefont {Barley}, \citenamefont {Bartnik}, \citenamefont {Bazarov},
  \citenamefont {Cultrera}, \citenamefont {Dobbins}, \citenamefont
  {Hoffstaetter}, \citenamefont {Johnson}, \citenamefont {Kaplan},
  \citenamefont {Karkare}, \citenamefont {Kostroun}, \citenamefont {Li},
  \citenamefont {Liepe}, \citenamefont {Liu}, \citenamefont {Loehl} \emph
  {et~al.}}]{Dunham2013}%
  \BibitemOpen
  \bibfield  {author} {\bibinfo {author} {\bibfnamefont {B.}~\bibnamefont
  {Dunham}}, \bibinfo {author} {\bibfnamefont {J.}~\bibnamefont {Barley}},
  \bibinfo {author} {\bibfnamefont {A.}~\bibnamefont {Bartnik}}, \bibinfo
  {author} {\bibfnamefont {I.}~\bibnamefont {Bazarov}}, \bibinfo {author}
  {\bibfnamefont {L.}~\bibnamefont {Cultrera}}, \bibinfo {author}
  {\bibfnamefont {J.}~\bibnamefont {Dobbins}}, \bibinfo {author} {\bibfnamefont
  {G.}~\bibnamefont {Hoffstaetter}}, \bibinfo {author} {\bibfnamefont
  {B.}~\bibnamefont {Johnson}}, \bibinfo {author} {\bibfnamefont
  {R.}~\bibnamefont {Kaplan}}, \bibinfo {author} {\bibfnamefont
  {S.}~\bibnamefont {Karkare}}, \bibinfo {author} {\bibfnamefont
  {V.}~\bibnamefont {Kostroun}}, \bibinfo {author} {\bibfnamefont
  {Y.}~\bibnamefont {Li}}, \bibinfo {author} {\bibfnamefont {M.}~\bibnamefont
  {Liepe}}, \bibinfo {author} {\bibfnamefont {X.}~\bibnamefont {Liu}}, \bibinfo
  {author} {\bibfnamefont {F.}~\bibnamefont {Loehl}}, \emph {et~al.},\ }\href
  {https://doi.org/10.1063/1.4789395} {\bibfield  {journal} {\bibinfo
  {journal} {Appl. Phys. Lett.}\ }\textbf {\bibinfo {volume} {102}},\ \bibinfo
  {pages} {034105} (\bibinfo {year} {2013})}\BibitemShut {NoStop}%
\bibitem [{\citenamefont {Orlov}\ \emph {et~al.}(2004)\citenamefont {Orlov},
  \citenamefont {Weigel}, \citenamefont {Schwalm}, \citenamefont {Terekhov},\
  and\ \citenamefont {Wolf}}]{Orlov2004}%
  \BibitemOpen
  \bibfield  {author} {\bibinfo {author} {\bibfnamefont {D.~A.}\ \bibnamefont
  {Orlov}}, \bibinfo {author} {\bibfnamefont {U.}~\bibnamefont {Weigel}},
  \bibinfo {author} {\bibfnamefont {D.}~\bibnamefont {Schwalm}}, \bibinfo
  {author} {\bibfnamefont {A.~S.}\ \bibnamefont {Terekhov}},\ and\ \bibinfo
  {author} {\bibfnamefont {A.}~\bibnamefont {Wolf}},\ }\href
  {https://doi.org/https://doi.org/10.1016/j.nima.2004.06.048} {\bibfield
  {journal} {\bibinfo  {journal} {Nucl. Instrum. Methods Phys. Res. A}\
  }\textbf {\bibinfo {volume} {532}},\ \bibinfo {pages} {418} (\bibinfo {year}
  {2004})}\BibitemShut {NoStop}%
\bibitem [{\citenamefont {Mamaev}\ \emph {et~al.}(2008)\citenamefont {Mamaev},
  \citenamefont {Gerchikov}, \citenamefont {Yashin}, \citenamefont {Vasiliev},
  \citenamefont {Kuzmichev}, \citenamefont {Ustinov}, \citenamefont {Zhukov},
  \citenamefont {Mikhrin},\ and\ \citenamefont {Vasiliev}}]{Mamaev2008}%
  \BibitemOpen
  \bibfield  {author} {\bibinfo {author} {\bibfnamefont {Y.~A.}\ \bibnamefont
  {Mamaev}}, \bibinfo {author} {\bibfnamefont {L.~G.}\ \bibnamefont
  {Gerchikov}}, \bibinfo {author} {\bibfnamefont {Y.~P.}\ \bibnamefont
  {Yashin}}, \bibinfo {author} {\bibfnamefont {D.~A.}\ \bibnamefont
  {Vasiliev}}, \bibinfo {author} {\bibfnamefont {V.~V.}\ \bibnamefont
  {Kuzmichev}}, \bibinfo {author} {\bibfnamefont {V.~M.}\ \bibnamefont
  {Ustinov}}, \bibinfo {author} {\bibfnamefont {A.~E.}\ \bibnamefont {Zhukov}},
  \bibinfo {author} {\bibfnamefont {V.~S.}\ \bibnamefont {Mikhrin}},\ and\
  \bibinfo {author} {\bibfnamefont {A.~P.}\ \bibnamefont {Vasiliev}},\ }\href
  {https://doi.org/10.1063/1.2976437} {\bibfield  {journal} {\bibinfo
  {journal} {Applied Physics Letters}\ }\textbf {\bibinfo {volume} {93}},\
  \bibinfo {pages} {081114} (\bibinfo {year} {2008})}\BibitemShut {NoStop}%
\bibitem [{\citenamefont {{Di Mitri}}\ and\ \citenamefont
  {Cornacchia}(2014)}]{DiMitri2014}%
  \BibitemOpen
  \bibfield  {author} {\bibinfo {author} {\bibfnamefont {S.}~\bibnamefont {{Di
  Mitri}}}\ and\ \bibinfo {author} {\bibfnamefont {M.}~\bibnamefont
  {Cornacchia}},\ }\href {https://doi.org/10.1016/j.physrep.2014.01.005}
  {\bibfield  {journal} {\bibinfo  {journal} {Phys. Rep.}\ }\textbf {\bibinfo
  {volume} {539}},\ \bibinfo {pages} {1} (\bibinfo {year} {2014})}\BibitemShut
  {NoStop}%
\bibitem [{\citenamefont {Novotný}\ \emph {et~al.}(2019)\citenamefont
  {Novotný}, \citenamefont {Wilhelm}, \citenamefont {Paul}, \citenamefont
  {Kálosi}, \citenamefont {Saurabh}, \citenamefont {Becker}, \citenamefont
  {Blaum}, \citenamefont {George}, \citenamefont {Göck}, \citenamefont
  {Grieser}, \citenamefont {Grussie}, \citenamefont {von Hahn}, \citenamefont
  {Krantz}, \citenamefont {Kreckel}, \citenamefont {Meyer} \emph
  {et~al.}}]{Novotny2019}%
  \BibitemOpen
  \bibfield  {author} {\bibinfo {author} {\bibfnamefont {O.}~\bibnamefont
  {Novotný}}, \bibinfo {author} {\bibfnamefont {P.}~\bibnamefont {Wilhelm}},
  \bibinfo {author} {\bibfnamefont {D.}~\bibnamefont {Paul}}, \bibinfo {author}
  {\bibfnamefont {A.}~\bibnamefont {Kálosi}}, \bibinfo {author} {\bibfnamefont
  {S.}~\bibnamefont {Saurabh}}, \bibinfo {author} {\bibfnamefont
  {A.}~\bibnamefont {Becker}}, \bibinfo {author} {\bibfnamefont
  {K.}~\bibnamefont {Blaum}}, \bibinfo {author} {\bibfnamefont
  {S.}~\bibnamefont {George}}, \bibinfo {author} {\bibfnamefont
  {J.}~\bibnamefont {Göck}}, \bibinfo {author} {\bibfnamefont
  {M.}~\bibnamefont {Grieser}}, \bibinfo {author} {\bibfnamefont
  {F.}~\bibnamefont {Grussie}}, \bibinfo {author} {\bibfnamefont
  {R.}~\bibnamefont {von Hahn}}, \bibinfo {author} {\bibfnamefont
  {C.}~\bibnamefont {Krantz}}, \bibinfo {author} {\bibfnamefont
  {H.}~\bibnamefont {Kreckel}}, \bibinfo {author} {\bibfnamefont
  {C.}~\bibnamefont {Meyer}}, \emph {et~al.},\ }\href
  {https://doi.org/10.1126/science.aax5921} {\bibfield  {journal} {\bibinfo
  {journal} {Science}\ }\textbf {\bibinfo {volume} {365}},\ \bibinfo {pages}
  {676} (\bibinfo {year} {2019})}\BibitemShut {NoStop}%
\bibitem [{\citenamefont {Filippetto}\ \emph {et~al.}(2022)\citenamefont
  {Filippetto}, \citenamefont {Musumeci}, \citenamefont {Li}, \citenamefont
  {Siwick}, \citenamefont {Otto}, \citenamefont {Centurion},\ and\
  \citenamefont {Nunes}}]{Filippetto2022}%
  \BibitemOpen
  \bibfield  {author} {\bibinfo {author} {\bibfnamefont {D.}~\bibnamefont
  {Filippetto}}, \bibinfo {author} {\bibfnamefont {P.}~\bibnamefont
  {Musumeci}}, \bibinfo {author} {\bibfnamefont {R.~K.}\ \bibnamefont {Li}},
  \bibinfo {author} {\bibfnamefont {B.~J.}\ \bibnamefont {Siwick}}, \bibinfo
  {author} {\bibfnamefont {M.~R.}\ \bibnamefont {Otto}}, \bibinfo {author}
  {\bibfnamefont {M.}~\bibnamefont {Centurion}},\ and\ \bibinfo {author}
  {\bibfnamefont {J.~P.~F.}\ \bibnamefont {Nunes}},\ }\href
  {https://doi.org/10.1103/RevModPhys.94.045004} {\bibfield  {journal}
  {\bibinfo  {journal} {Rev. Mod. Phys.}\ }\textbf {\bibinfo {volume} {94}},\
  \bibinfo {pages} {045004} (\bibinfo {year} {2022})}\BibitemShut {NoStop}%
\bibitem [{\citenamefont {Pomarico}\ and\ \citenamefont
  {Kim}(2018)}]{Pomarico2018}%
  \BibitemOpen
  \bibfield  {author} {\bibinfo {author} {\bibfnamefont {E.}~\bibnamefont
  {Pomarico}}\ and\ \bibinfo {author} {\bibfnamefont {Y.-J.}\ \bibnamefont
  {Kim}},\ }\href {https://doi.org/10.1557/mrs.2018.148} {\bibfield  {journal}
  {\bibinfo  {journal} {MRS Bulletin}\ }\textbf {\bibinfo {volume} {43}},\
  \bibinfo {pages} {497} (\bibinfo {year} {2018})}\BibitemShut {NoStop}%
\bibitem [{\citenamefont {Musumeci}\ \emph {et~al.}(2018)\citenamefont
  {Musumeci}, \citenamefont {Navarro}, \citenamefont {Rosenzweig},
  \citenamefont {Cultrera}, \citenamefont {Bazarov}, \citenamefont {Maxson},
  \citenamefont {Karkare},\ and\ \citenamefont {Padmore}}]{Musumeci2018}%
  \BibitemOpen
  \bibfield  {author} {\bibinfo {author} {\bibfnamefont {P.}~\bibnamefont
  {Musumeci}}, \bibinfo {author} {\bibfnamefont {J.~G.}\ \bibnamefont
  {Navarro}}, \bibinfo {author} {\bibfnamefont {J.~B.}\ \bibnamefont
  {Rosenzweig}}, \bibinfo {author} {\bibfnamefont {L.}~\bibnamefont
  {Cultrera}}, \bibinfo {author} {\bibfnamefont {I.}~\bibnamefont {Bazarov}},
  \bibinfo {author} {\bibfnamefont {J.}~\bibnamefont {Maxson}}, \bibinfo
  {author} {\bibfnamefont {S.}~\bibnamefont {Karkare}},\ and\ \bibinfo {author}
  {\bibfnamefont {H.}~\bibnamefont {Padmore}},\ }\href
  {https://doi.org/10.1016/j.nima.2018.03.019} {\bibfield  {journal} {\bibinfo
  {journal} {Nucl. Instrum. Methods Phys. Res. A}\ }\textbf {\bibinfo {volume}
  {907}},\ \bibinfo {pages} {209} (\bibinfo {year} {2018})}\BibitemShut
  {NoStop}%
\bibitem [{\citenamefont {Rodway}\ and\ \citenamefont
  {Allenson}(1986)}]{Rodway1986}%
  \BibitemOpen
  \bibfield  {author} {\bibinfo {author} {\bibfnamefont {D.~C.}\ \bibnamefont
  {Rodway}}\ and\ \bibinfo {author} {\bibfnamefont {M.~B.}\ \bibnamefont
  {Allenson}},\ }\href {https://doi.org/10.1088/0022-3727/19/7/024} {\bibfield
  {journal} {\bibinfo  {journal} {J. Phys. D: Appl. Phys.}\ }\textbf {\bibinfo
  {volume} {19}},\ \bibinfo {pages} {1353} (\bibinfo {year}
  {1986})}\BibitemShut {NoStop}%
\bibitem [{\citenamefont {Karkare}\ \emph {et~al.}(2014)\citenamefont
  {Karkare}, \citenamefont {Boulet}, \citenamefont {Cultrera}, \citenamefont
  {Dunham}, \citenamefont {Liu}, \citenamefont {Schaff},\ and\ \citenamefont
  {Bazarov}}]{Karkare2014}%
  \BibitemOpen
  \bibfield  {author} {\bibinfo {author} {\bibfnamefont {S.}~\bibnamefont
  {Karkare}}, \bibinfo {author} {\bibfnamefont {L.}~\bibnamefont {Boulet}},
  \bibinfo {author} {\bibfnamefont {L.}~\bibnamefont {Cultrera}}, \bibinfo
  {author} {\bibfnamefont {B.}~\bibnamefont {Dunham}}, \bibinfo {author}
  {\bibfnamefont {X.}~\bibnamefont {Liu}}, \bibinfo {author} {\bibfnamefont
  {W.}~\bibnamefont {Schaff}},\ and\ \bibinfo {author} {\bibfnamefont
  {I.}~\bibnamefont {Bazarov}},\ }\href
  {https://doi.org/10.1103/PhysRevLett.112.097601} {\bibfield  {journal}
  {\bibinfo  {journal} {Phys. Rev. Lett.}\ }\textbf {\bibinfo {volume} {112}},\
  \bibinfo {pages} {097601} (\bibinfo {year} {2014})}\BibitemShut {NoStop}%
\bibitem [{\citenamefont {Feng}\ \emph {et~al.}(2015)\citenamefont {Feng},
  \citenamefont {Nasiatka}, \citenamefont {Wan}, \citenamefont {Karkare},
  \citenamefont {Smedley},\ and\ \citenamefont {Padmore}}]{Feng2015}%
  \BibitemOpen
  \bibfield  {author} {\bibinfo {author} {\bibfnamefont {J.}~\bibnamefont
  {Feng}}, \bibinfo {author} {\bibfnamefont {J.}~\bibnamefont {Nasiatka}},
  \bibinfo {author} {\bibfnamefont {W.}~\bibnamefont {Wan}}, \bibinfo {author}
  {\bibfnamefont {S.}~\bibnamefont {Karkare}}, \bibinfo {author} {\bibfnamefont
  {J.}~\bibnamefont {Smedley}},\ and\ \bibinfo {author} {\bibfnamefont {H.~A.}\
  \bibnamefont {Padmore}},\ }\href {https://doi.org/10.1063/1.4931976}
  {\bibfield  {journal} {\bibinfo  {journal} {Appl. Phys. Lett.}\ }\textbf
  {\bibinfo {volume} {107}},\ \bibinfo {pages} {134101} (\bibinfo {year}
  {2015})}\BibitemShut {NoStop}%
\bibitem [{\citenamefont {Karkare}\ \emph {et~al.}(2020)\citenamefont
  {Karkare}, \citenamefont {Adhikari}, \citenamefont {Schroeder}, \citenamefont
  {Nangoi}, \citenamefont {Arias}, \citenamefont {Maxson},\ and\ \citenamefont
  {Padmore}}]{Karkare2020}%
  \BibitemOpen
  \bibfield  {author} {\bibinfo {author} {\bibfnamefont {S.}~\bibnamefont
  {Karkare}}, \bibinfo {author} {\bibfnamefont {G.}~\bibnamefont {Adhikari}},
  \bibinfo {author} {\bibfnamefont {W.~A.}\ \bibnamefont {Schroeder}}, \bibinfo
  {author} {\bibfnamefont {J.~K.}\ \bibnamefont {Nangoi}}, \bibinfo {author}
  {\bibfnamefont {T.}~\bibnamefont {Arias}}, \bibinfo {author} {\bibfnamefont
  {J.}~\bibnamefont {Maxson}},\ and\ \bibinfo {author} {\bibfnamefont
  {H.}~\bibnamefont {Padmore}},\ }\href
  {https://doi.org/10.1103/PhysRevLett.125.054801} {\bibfield  {journal}
  {\bibinfo  {journal} {Phys. Rev. Lett.}\ }\textbf {\bibinfo {volume} {125}},\
  \bibinfo {pages} {054801} (\bibinfo {year} {2020})}\BibitemShut {NoStop}%
\bibitem [{\citenamefont {Nangoi}\ \emph {et~al.}(2021)\citenamefont {Nangoi},
  \citenamefont {Karkare}, \citenamefont {Sundararaman}, \citenamefont
  {Padmore},\ and\ \citenamefont {Arias}}]{Nangoi2021}%
  \BibitemOpen
  \bibfield  {author} {\bibinfo {author} {\bibfnamefont {J.~K.}\ \bibnamefont
  {Nangoi}}, \bibinfo {author} {\bibfnamefont {S.}~\bibnamefont {Karkare}},
  \bibinfo {author} {\bibfnamefont {R.}~\bibnamefont {Sundararaman}}, \bibinfo
  {author} {\bibfnamefont {H.~A.}\ \bibnamefont {Padmore}},\ and\ \bibinfo
  {author} {\bibfnamefont {T.~A.}\ \bibnamefont {Arias}},\ }\href
  {https://doi.org/10.1103/PhysRevB.104.115132} {\bibfield  {journal} {\bibinfo
   {journal} {Phys. Rev. B}\ }\textbf {\bibinfo {volume} {104}},\ \bibinfo
  {pages} {115132} (\bibinfo {year} {2021})}\BibitemShut {NoStop}%
\bibitem [{\citenamefont {Schroeder}\ \emph {et~al.}(2025)\citenamefont
  {Schroeder}, \citenamefont {Angeloni}, \citenamefont {Shan},\ and\
  \citenamefont {Jones}}]{Schroeder2025}%
  \BibitemOpen
  \bibfield  {author} {\bibinfo {author} {\bibfnamefont {W.~A.}\ \bibnamefont
  {Schroeder}}, \bibinfo {author} {\bibfnamefont {L.~A.}\ \bibnamefont
  {Angeloni}}, \bibinfo {author} {\bibfnamefont {I.-J.}\ \bibnamefont {Shan}},\
  and\ \bibinfo {author} {\bibfnamefont {L.~B.}\ \bibnamefont {Jones}},\ }\href
  {https://doi.org/10.1103/PhysRevApplied.23.054065} {\bibfield  {journal}
  {\bibinfo  {journal} {Phys. Rev. Appl.}\ }\textbf {\bibinfo {volume} {23}},\
  \bibinfo {pages} {054065} (\bibinfo {year} {2025})}\BibitemShut {NoStop}%
\bibitem [{\citenamefont {Vergara}\ \emph {et~al.}(1996)\citenamefont
  {Vergara}, \citenamefont {Herrera‐Gómez},\ and\ \citenamefont
  {Spicer}}]{Vergara1996}%
  \BibitemOpen
  \bibfield  {author} {\bibinfo {author} {\bibfnamefont {G.}~\bibnamefont
  {Vergara}}, \bibinfo {author} {\bibfnamefont {A.}~\bibnamefont
  {Herrera‐Gómez}},\ and\ \bibinfo {author} {\bibfnamefont {W.~E.}\
  \bibnamefont {Spicer}},\ }\href {https://doi.org/10.1063/1.362992} {\bibfield
   {journal} {\bibinfo  {journal} {J. Appl. Phys.}\ }\textbf {\bibinfo {volume}
  {80}},\ \bibinfo {pages} {1809} (\bibinfo {year} {1996})}\BibitemShut
  {NoStop}%
\bibitem [{\citenamefont {Rickman}\ \emph {et~al.}(2013)\citenamefont
  {Rickman}, \citenamefont {Berger}, \citenamefont {Nicholls},\ and\
  \citenamefont {Schroeder}}]{Rickman2013}%
  \BibitemOpen
  \bibfield  {author} {\bibinfo {author} {\bibfnamefont {B.~L.}\ \bibnamefont
  {Rickman}}, \bibinfo {author} {\bibfnamefont {J.~A.}\ \bibnamefont {Berger}},
  \bibinfo {author} {\bibfnamefont {A.~W.}\ \bibnamefont {Nicholls}},\ and\
  \bibinfo {author} {\bibfnamefont {W.~A.}\ \bibnamefont {Schroeder}},\ }\href
  {https://doi.org/10.1103/PhysRevLett.111.237401} {\bibfield  {journal}
  {\bibinfo  {journal} {Phys. Rev. Lett.}\ }\textbf {\bibinfo {volume} {111}},\
  \bibinfo {pages} {237401} (\bibinfo {year} {2013})}\BibitemShut {NoStop}%
\bibitem [{\citenamefont {Rickman}\ \emph {et~al.}(2014)\citenamefont
  {Rickman}, \citenamefont {Berger}, \citenamefont {Nicholls},\ and\
  \citenamefont {Schroeder}}]{Rickman2014}%
  \BibitemOpen
  \bibfield  {author} {\bibinfo {author} {\bibfnamefont {B.~L.}\ \bibnamefont
  {Rickman}}, \bibinfo {author} {\bibfnamefont {J.~A.}\ \bibnamefont {Berger}},
  \bibinfo {author} {\bibfnamefont {A.~W.}\ \bibnamefont {Nicholls}},\ and\
  \bibinfo {author} {\bibfnamefont {W.~A.}\ \bibnamefont {Schroeder}},\ }\href
  {https://doi.org/10.1103/PhysRevLett.113.239904} {\bibfield  {journal}
  {\bibinfo  {journal} {Phys. Rev. Lett.}\ }\textbf {\bibinfo {volume} {113}},\
  \bibinfo {pages} {239904} (\bibinfo {year} {2014})}\BibitemShut {NoStop}%
\bibitem [{\citenamefont {Alperovich}\ \emph {et~al.}(2021)\citenamefont
  {Alperovich}, \citenamefont {Kazantsev}, \citenamefont {Zhuravlev},\ and\
  \citenamefont {Shvartsman}}]{Alperovich2021}%
  \BibitemOpen
  \bibfield  {author} {\bibinfo {author} {\bibfnamefont {V.~L.}\ \bibnamefont
  {Alperovich}}, \bibinfo {author} {\bibfnamefont {D.~M.}\ \bibnamefont
  {Kazantsev}}, \bibinfo {author} {\bibfnamefont {A.~G.}\ \bibnamefont
  {Zhuravlev}},\ and\ \bibinfo {author} {\bibfnamefont {L.~D.}\ \bibnamefont
  {Shvartsman}},\ }\href {https://doi.org/10.1016/j.apsusc.2021.149987}
  {\bibfield  {journal} {\bibinfo  {journal} {Appl. Surf. Sci.}\ }\textbf
  {\bibinfo {volume} {561}},\ \bibinfo {pages} {149987} (\bibinfo {year}
  {2021})}\BibitemShut {NoStop}%
\bibitem [{\citenamefont {Pollard}(1974)}]{Pollard1974}%
  \BibitemOpen
  \bibfield  {author} {\bibinfo {author} {\bibfnamefont {J.~H.}\ \bibnamefont
  {Pollard}},\ }in\ \href@noop {} {\emph {\bibinfo {booktitle} {Proc. 2nd
  {E}uropean {E}lectro-Optics {M}arkets and {T}echnology {C}onference}}}\
  (\bibinfo {year} {1974})\ p.\ \bibinfo {pages} {316}\BibitemShut {NoStop}%
\bibitem [{\citenamefont {Lee}\ \emph {et~al.}(2007)\citenamefont {Lee},
  \citenamefont {Sun}, \citenamefont {Liu}, \citenamefont {Sun},\ and\
  \citenamefont {Pianetta}}]{Lee2007}%
  \BibitemOpen
  \bibfield  {author} {\bibinfo {author} {\bibfnamefont {D.-I.}\ \bibnamefont
  {Lee}}, \bibinfo {author} {\bibfnamefont {Y.}~\bibnamefont {Sun}}, \bibinfo
  {author} {\bibfnamefont {Z.}~\bibnamefont {Liu}}, \bibinfo {author}
  {\bibfnamefont {S.}~\bibnamefont {Sun}},\ and\ \bibinfo {author}
  {\bibfnamefont {P.}~\bibnamefont {Pianetta}},\ }\href
  {https://doi.org/10.1063/1.2805775} {\bibfield  {journal} {\bibinfo
  {journal} {Appl. Phys. Lett.}\ }\textbf {\bibinfo {volume} {91}},\ \bibinfo
  {pages} {192101} (\bibinfo {year} {2007})}\BibitemShut {NoStop}%
\bibitem [{\citenamefont {Bradley}\ \emph {et~al.}(1977)\citenamefont
  {Bradley}, \citenamefont {Allenson},\ and\ \citenamefont
  {Holeman}}]{Bradley1977}%
  \BibitemOpen
  \bibfield  {author} {\bibinfo {author} {\bibfnamefont {D.~J.}\ \bibnamefont
  {Bradley}}, \bibinfo {author} {\bibfnamefont {M.~B.}\ \bibnamefont
  {Allenson}},\ and\ \bibinfo {author} {\bibfnamefont {B.~R.}\ \bibnamefont
  {Holeman}},\ }\href {https://doi.org/10.1088/0022-3727/10/1/013} {\bibfield
  {journal} {\bibinfo  {journal} {J. Phys. D: Appl. Phys.}\ }\textbf {\bibinfo
  {volume} {10}},\ \bibinfo {pages} {111} (\bibinfo {year} {1977})}\BibitemShut
  {NoStop}%
\bibitem [{\citenamefont {Lee}\ \emph {et~al.}(2015)\citenamefont {Lee},
  \citenamefont {Karkare}, \citenamefont {Cultrera}, \citenamefont {Kim},\ and\
  \citenamefont {Bazarov}}]{Lee2015}%
  \BibitemOpen
  \bibfield  {author} {\bibinfo {author} {\bibfnamefont {H.}~\bibnamefont
  {Lee}}, \bibinfo {author} {\bibfnamefont {S.}~\bibnamefont {Karkare}},
  \bibinfo {author} {\bibfnamefont {L.}~\bibnamefont {Cultrera}}, \bibinfo
  {author} {\bibfnamefont {A.}~\bibnamefont {Kim}},\ and\ \bibinfo {author}
  {\bibfnamefont {I.~V.}\ \bibnamefont {Bazarov}},\ }\href
  {https://doi.org/10.1063/1.4927381} {\bibfield  {journal} {\bibinfo
  {journal} {Rev. Sci. Instrum.}\ }\textbf {\bibinfo {volume} {86}},\ \bibinfo
  {pages} {073309} (\bibinfo {year} {2015})}\BibitemShut {NoStop}%
\bibitem [{\citenamefont {Orlov}\ \emph {et~al.}(2001)\citenamefont {Orlov},
  \citenamefont {Hoppe}, \citenamefont {Weigel}, \citenamefont {Schwalm},
  \citenamefont {Terekhov},\ and\ \citenamefont {Wolf}}]{Orlov2001}%
  \BibitemOpen
  \bibfield  {author} {\bibinfo {author} {\bibfnamefont {D.~A.}\ \bibnamefont
  {Orlov}}, \bibinfo {author} {\bibfnamefont {M.}~\bibnamefont {Hoppe}},
  \bibinfo {author} {\bibfnamefont {U.}~\bibnamefont {Weigel}}, \bibinfo
  {author} {\bibfnamefont {D.}~\bibnamefont {Schwalm}}, \bibinfo {author}
  {\bibfnamefont {A.~S.}\ \bibnamefont {Terekhov}},\ and\ \bibinfo {author}
  {\bibfnamefont {A.}~\bibnamefont {Wolf}},\ }\href
  {https://doi.org/10.1063/1.1368376} {\bibfield  {journal} {\bibinfo
  {journal} {Appl. Phys. Lett.}\ }\textbf {\bibinfo {volume} {78}},\ \bibinfo
  {pages} {2721} (\bibinfo {year} {2001})}\BibitemShut {NoStop}%
\bibitem [{\citenamefont {Bakin}\ \emph {et~al.}(2003)\citenamefont {Bakin},
  \citenamefont {Pakhnevich}, \citenamefont {Kosolobov}, \citenamefont
  {Scheibler}, \citenamefont {Jaroshevich},\ and\ \citenamefont
  {Terekhov}}]{Bakin2003}%
  \BibitemOpen
  \bibfield  {author} {\bibinfo {author} {\bibfnamefont {V.~V.}\ \bibnamefont
  {Bakin}}, \bibinfo {author} {\bibfnamefont {A.~A.}\ \bibnamefont
  {Pakhnevich}}, \bibinfo {author} {\bibfnamefont {S.~N.}\ \bibnamefont
  {Kosolobov}}, \bibinfo {author} {\bibfnamefont {H.~E.}\ \bibnamefont
  {Scheibler}}, \bibinfo {author} {\bibfnamefont {A.~S.}\ \bibnamefont
  {Jaroshevich}},\ and\ \bibinfo {author} {\bibfnamefont {A.~S.}\ \bibnamefont
  {Terekhov}},\ }\href {https://doi.org/10.1134/1.1571875} {\bibfield
  {journal} {\bibinfo  {journal} {JETP Lett.}\ }\textbf {\bibinfo {volume}
  {77}},\ \bibinfo {pages} {167} (\bibinfo {year} {2003})}\BibitemShut
  {NoStop}%
\bibitem [{\citenamefont {Karkare}\ and\ \citenamefont
  {Bazarov}(2011)}]{Karkare2011}%
  \BibitemOpen
  \bibfield  {author} {\bibinfo {author} {\bibfnamefont {S.}~\bibnamefont
  {Karkare}}\ and\ \bibinfo {author} {\bibfnamefont {I.}~\bibnamefont
  {Bazarov}},\ }\href {https://doi.org/10.1063/1.3559895} {\bibfield  {journal}
  {\bibinfo  {journal} {Appl. Phys. Lett.}\ }\textbf {\bibinfo {volume} {98}},\
  \bibinfo {pages} {094104} (\bibinfo {year} {2011})}\BibitemShut {NoStop}%
\bibitem [{\citenamefont {Jones}\ \emph {et~al.}(2021)\citenamefont {Jones},
  \citenamefont {Scheibler}, \citenamefont {Kosolobov}, \citenamefont
  {Terekhov}, \citenamefont {Militsyn},\ and\ \citenamefont
  {Noakes}}]{Jones2021}%
  \BibitemOpen
  \bibfield  {author} {\bibinfo {author} {\bibfnamefont {L.~B.}\ \bibnamefont
  {Jones}}, \bibinfo {author} {\bibfnamefont {H.~E.}\ \bibnamefont
  {Scheibler}}, \bibinfo {author} {\bibfnamefont {S.~N.}\ \bibnamefont
  {Kosolobov}}, \bibinfo {author} {\bibfnamefont {A.~S.}\ \bibnamefont
  {Terekhov}}, \bibinfo {author} {\bibfnamefont {B.~L.}\ \bibnamefont
  {Militsyn}},\ and\ \bibinfo {author} {\bibfnamefont {T.~C.~Q.}\ \bibnamefont
  {Noakes}},\ }\href {https://doi.org/10.1088/1361-6463/abe1e9} {\bibfield
  {journal} {\bibinfo  {journal} {J. Phys. D: Appl. Phys.}\ }\textbf {\bibinfo
  {volume} {54}},\ \bibinfo {pages} {205301} (\bibinfo {year}
  {2021})}\BibitemShut {NoStop}%
\bibitem [{\citenamefont {Karkare}\ \emph {et~al.}(2017)\citenamefont
  {Karkare}, \citenamefont {Feng}, \citenamefont {Chen}, \citenamefont {Wan},
  \citenamefont {Palomares}, \citenamefont {Chiang},\ and\ \citenamefont
  {Padmore}}]{Karkare2017}%
  \BibitemOpen
  \bibfield  {author} {\bibinfo {author} {\bibfnamefont {S.}~\bibnamefont
  {Karkare}}, \bibinfo {author} {\bibfnamefont {J.}~\bibnamefont {Feng}},
  \bibinfo {author} {\bibfnamefont {X.}~\bibnamefont {Chen}}, \bibinfo {author}
  {\bibfnamefont {W.}~\bibnamefont {Wan}}, \bibinfo {author} {\bibfnamefont
  {F.~J.}\ \bibnamefont {Palomares}}, \bibinfo {author} {\bibfnamefont {T.-C.}\
  \bibnamefont {Chiang}},\ and\ \bibinfo {author} {\bibfnamefont {H.~A.}\
  \bibnamefont {Padmore}},\ }\href
  {https://doi.org/10.1103/PhysRevLett.118.164802} {\bibfield  {journal}
  {\bibinfo  {journal} {Phys. Rev. Lett.}\ }\textbf {\bibinfo {volume} {118}},\
  \bibinfo {pages} {164802} (\bibinfo {year} {2017})}\BibitemShut {NoStop}%
\bibitem [{\citenamefont {Goldstein}(1975)}]{Goldstein1975}%
  \BibitemOpen
  \bibfield  {author} {\bibinfo {author} {\bibfnamefont {B.}~\bibnamefont
  {Goldstein}},\ }\href {https://doi.org/10.1016/0039-6028(75)90280-0}
  {\bibfield  {journal} {\bibinfo  {journal} {Surf. Sci.}\ }\textbf {\bibinfo
  {volume} {47}},\ \bibinfo {pages} {143} (\bibinfo {year} {1975})}\BibitemShut
  {NoStop}%
\bibitem [{\citenamefont {Bakin}\ \emph {et~al.}(2007)\citenamefont {Bakin},
  \citenamefont {Pakhnevich}, \citenamefont {Zhuravlev}, \citenamefont
  {Shornikov}, \citenamefont {Akhundov}, \citenamefont {Tereshechenko},
  \citenamefont {Alperovich}, \citenamefont {Scheibler},\ and\ \citenamefont
  {Terekhov}}]{Bakin2007}%
  \BibitemOpen
  \bibfield  {author} {\bibinfo {author} {\bibfnamefont {V.~V.}\ \bibnamefont
  {Bakin}}, \bibinfo {author} {\bibfnamefont {A.~A.}\ \bibnamefont
  {Pakhnevich}}, \bibinfo {author} {\bibfnamefont {A.~G.}\ \bibnamefont
  {Zhuravlev}}, \bibinfo {author} {\bibfnamefont {A.~N.}\ \bibnamefont
  {Shornikov}}, \bibinfo {author} {\bibfnamefont {I.~O.}\ \bibnamefont
  {Akhundov}}, \bibinfo {author} {\bibfnamefont {O.~E.}\ \bibnamefont
  {Tereshechenko}}, \bibinfo {author} {\bibfnamefont {V.~L.}\ \bibnamefont
  {Alperovich}}, \bibinfo {author} {\bibfnamefont {H.~E.}\ \bibnamefont
  {Scheibler}},\ and\ \bibinfo {author} {\bibfnamefont {A.~S.}\ \bibnamefont
  {Terekhov}},\ }\href {https://doi.org/10.1380/ejssnt.2007.80} {\bibfield
  {journal} {\bibinfo  {journal} {e-J. Surf. Sci. Nanotechnol.}\ }\textbf
  {\bibinfo {volume} {5}},\ \bibinfo {pages} {80} (\bibinfo {year}
  {2007})}\BibitemShut {NoStop}%
\bibitem [{\citenamefont {Berger}\ \emph {et~al.}(2012)\citenamefont {Berger},
  \citenamefont {Rickman}, \citenamefont {Li}, \citenamefont {Nicholls},\ and\
  \citenamefont {Andreas~Schroeder}}]{Berger2012}%
  \BibitemOpen
  \bibfield  {author} {\bibinfo {author} {\bibfnamefont {J.~A.}\ \bibnamefont
  {Berger}}, \bibinfo {author} {\bibfnamefont {B.~L.}\ \bibnamefont {Rickman}},
  \bibinfo {author} {\bibfnamefont {T.}~\bibnamefont {Li}}, \bibinfo {author}
  {\bibfnamefont {A.~W.}\ \bibnamefont {Nicholls}},\ and\ \bibinfo {author}
  {\bibfnamefont {W.}~\bibnamefont {Andreas~Schroeder}},\ }\href
  {https://doi.org/10.1063/1.4766350} {\bibfield  {journal} {\bibinfo
  {journal} {Appl. Phys. Lett.}\ }\textbf {\bibinfo {volume} {101}},\ \bibinfo
  {pages} {194103} (\bibinfo {year} {2012})}\BibitemShut {NoStop}%
\bibitem [{\citenamefont {Li}\ \emph {et~al.}(2015)\citenamefont {Li},
  \citenamefont {Rickman},\ and\ \citenamefont {Schroeder}}]{Li2015}%
  \BibitemOpen
  \bibfield  {author} {\bibinfo {author} {\bibfnamefont {T.}~\bibnamefont
  {Li}}, \bibinfo {author} {\bibfnamefont {B.~L.}\ \bibnamefont {Rickman}},\
  and\ \bibinfo {author} {\bibfnamefont {W.~A.}\ \bibnamefont {Schroeder}},\
  }\href {https://doi.org/10.1063/1.4916598} {\bibfield  {journal} {\bibinfo
  {journal} {J. Appl. Phys.}\ }\textbf {\bibinfo {volume} {117}},\ \bibinfo
  {pages} {134901} (\bibinfo {year} {2015})}\BibitemShut {NoStop}%
\bibitem [{\citenamefont {Dowell}\ and\ \citenamefont
  {Schmerge}(2009)}]{Dowell2009}%
  \BibitemOpen
  \bibfield  {author} {\bibinfo {author} {\bibfnamefont {D.~H.}\ \bibnamefont
  {Dowell}}\ and\ \bibinfo {author} {\bibfnamefont {J.~F.}\ \bibnamefont
  {Schmerge}},\ }\href {https://doi.org/10.1103/PhysRevSTAB.12.074201}
  {\bibfield  {journal} {\bibinfo  {journal} {Phys. Rev. ST Accel. Beams}\
  }\textbf {\bibinfo {volume} {12}},\ \bibinfo {pages} {074201} (\bibinfo
  {year} {2009})}\BibitemShut {NoStop}%
\bibitem [{\citenamefont {Spicer}(1958)}]{Spicer1958}%
  \BibitemOpen
  \bibfield  {author} {\bibinfo {author} {\bibfnamefont {W.~E.}\ \bibnamefont
  {Spicer}},\ }\href {https://doi.org/10.1103/PhysRev.112.114} {\bibfield
  {journal} {\bibinfo  {journal} {Phys. Rev.}\ }\textbf {\bibinfo {volume}
  {112}},\ \bibinfo {pages} {114} (\bibinfo {year} {1958})}\BibitemShut
  {NoStop}%
\bibitem [{\citenamefont {Petrushina}\ \emph {et~al.}(2020)\citenamefont
  {Petrushina}, \citenamefont {Litvinenko}, \citenamefont {Jing}, \citenamefont
  {Ma}, \citenamefont {Pinayev}, \citenamefont {Shih}, \citenamefont {Wang},
  \citenamefont {Wu}, \citenamefont {Altinbas}, \citenamefont {Brutus},
  \citenamefont {Belomestnykh}, \citenamefont {Di~Lieto}, \citenamefont
  {Inacker}, \citenamefont {Jamilkowski}, \citenamefont {Mahler} \emph
  {et~al.}}]{Petrushina2020}%
  \BibitemOpen
  \bibfield  {author} {\bibinfo {author} {\bibfnamefont {I.}~\bibnamefont
  {Petrushina}}, \bibinfo {author} {\bibfnamefont {V.~N.}\ \bibnamefont
  {Litvinenko}}, \bibinfo {author} {\bibfnamefont {Y.}~\bibnamefont {Jing}},
  \bibinfo {author} {\bibfnamefont {J.}~\bibnamefont {Ma}}, \bibinfo {author}
  {\bibfnamefont {I.}~\bibnamefont {Pinayev}}, \bibinfo {author} {\bibfnamefont
  {K.}~\bibnamefont {Shih}}, \bibinfo {author} {\bibfnamefont {G.}~\bibnamefont
  {Wang}}, \bibinfo {author} {\bibfnamefont {Y.~H.}\ \bibnamefont {Wu}},
  \bibinfo {author} {\bibfnamefont {Z.}~\bibnamefont {Altinbas}}, \bibinfo
  {author} {\bibfnamefont {J.~C.}\ \bibnamefont {Brutus}}, \bibinfo {author}
  {\bibfnamefont {S.}~\bibnamefont {Belomestnykh}}, \bibinfo {author}
  {\bibfnamefont {A.}~\bibnamefont {Di~Lieto}}, \bibinfo {author}
  {\bibfnamefont {P.}~\bibnamefont {Inacker}}, \bibinfo {author} {\bibfnamefont
  {J.}~\bibnamefont {Jamilkowski}}, \bibinfo {author} {\bibfnamefont
  {G.}~\bibnamefont {Mahler}}, \emph {et~al.},\ }\href
  {https://doi.org/10.1103/PhysRevLett.124.244801} {\bibfield  {journal}
  {\bibinfo  {journal} {Phys. Rev. Lett.}\ }\textbf {\bibinfo {volume} {124}},\
  \bibinfo {pages} {244801} (\bibinfo {year} {2020})}\BibitemShut {NoStop}%
\bibitem [{\citenamefont {Cultrera}\ \emph {et~al.}(2016)\citenamefont
  {Cultrera}, \citenamefont {Gulliford}, \citenamefont {Bartnik}, \citenamefont
  {Lee},\ and\ \citenamefont {Bazarov}}]{Cultrera2016}%
  \BibitemOpen
  \bibfield  {author} {\bibinfo {author} {\bibfnamefont {L.}~\bibnamefont
  {Cultrera}}, \bibinfo {author} {\bibfnamefont {C.}~\bibnamefont {Gulliford}},
  \bibinfo {author} {\bibfnamefont {A.}~\bibnamefont {Bartnik}}, \bibinfo
  {author} {\bibfnamefont {H.}~\bibnamefont {Lee}},\ and\ \bibinfo {author}
  {\bibfnamefont {I.}~\bibnamefont {Bazarov}},\ }\href
  {https://doi.org/10.1063/1.4945091} {\bibfield  {journal} {\bibinfo
  {journal} {Appl. Phys. Lett.}\ }\textbf {\bibinfo {volume} {108}},\ \bibinfo
  {pages} {134105} (\bibinfo {year} {2016})}\BibitemShut {NoStop}%
\bibitem [{\citenamefont {Dube}\ \emph {et~al.}(2025)\citenamefont {Dube},
  \citenamefont {Kühn}, \citenamefont {Wang}, \citenamefont {Mistry},
  \citenamefont {Klemz}, \citenamefont {Galdi},\ and\ \citenamefont
  {Kamps}}]{Dube2025}%
  \BibitemOpen
  \bibfield  {author} {\bibinfo {author} {\bibfnamefont {J.}~\bibnamefont
  {Dube}}, \bibinfo {author} {\bibfnamefont {J.}~\bibnamefont {Kühn}},
  \bibinfo {author} {\bibfnamefont {C.}~\bibnamefont {Wang}}, \bibinfo {author}
  {\bibfnamefont {S.}~\bibnamefont {Mistry}}, \bibinfo {author} {\bibfnamefont
  {G.}~\bibnamefont {Klemz}}, \bibinfo {author} {\bibfnamefont
  {A.}~\bibnamefont {Galdi}},\ and\ \bibinfo {author} {\bibfnamefont
  {T.}~\bibnamefont {Kamps}},\ }\href {https://doi.org/10.1063/5.0268697}
  {\bibfield  {journal} {\bibinfo  {journal} {J. Appl. Phys.}\ }\textbf
  {\bibinfo {volume} {138}},\ \bibinfo {pages} {045704} (\bibinfo {year}
  {2025})}\BibitemShut {NoStop}%
\bibitem [{\citenamefont {Rusetsky}\ \emph {et~al.}(2022)\citenamefont
  {Rusetsky}, \citenamefont {Golyashov}, \citenamefont {Eremeev}, \citenamefont
  {Kustov}, \citenamefont {Rusinov}, \citenamefont {Shamirzaev}, \citenamefont
  {Mironov}, \citenamefont {Demin},\ and\ \citenamefont
  {Tereshchenko}}]{Rusetsky2022}%
  \BibitemOpen
  \bibfield  {author} {\bibinfo {author} {\bibfnamefont {V.~S.}\ \bibnamefont
  {Rusetsky}}, \bibinfo {author} {\bibfnamefont {V.~A.}\ \bibnamefont
  {Golyashov}}, \bibinfo {author} {\bibfnamefont {S.~V.}\ \bibnamefont
  {Eremeev}}, \bibinfo {author} {\bibfnamefont {D.~A.}\ \bibnamefont {Kustov}},
  \bibinfo {author} {\bibfnamefont {I.~P.}\ \bibnamefont {Rusinov}}, \bibinfo
  {author} {\bibfnamefont {T.~S.}\ \bibnamefont {Shamirzaev}}, \bibinfo
  {author} {\bibfnamefont {A.~V.}\ \bibnamefont {Mironov}}, \bibinfo {author}
  {\bibfnamefont {A.~Y.}\ \bibnamefont {Demin}},\ and\ \bibinfo {author}
  {\bibfnamefont {O.~E.}\ \bibnamefont {Tereshchenko}},\ }\href
  {https://doi.org/10.1103/PhysRevLett.129.166802} {\bibfield  {journal}
  {\bibinfo  {journal} {Phys. Rev. Lett.}\ }\textbf {\bibinfo {volume} {129}},\
  \bibinfo {pages} {166802} (\bibinfo {year} {2022})}\BibitemShut {NoStop}%
\bibitem [{\citenamefont {Ninomiya}\ \emph {et~al.}(1969)\citenamefont
  {Ninomiya}, \citenamefont {Taketoshi},\ and\ \citenamefont
  {Tachiya}}]{Ninomiya1969}%
  \BibitemOpen
  \bibfield  {author} {\bibinfo {author} {\bibfnamefont {T.}~\bibnamefont
  {Ninomiya}}, \bibinfo {author} {\bibfnamefont {K.}~\bibnamefont
  {Taketoshi}},\ and\ \bibinfo {author} {\bibfnamefont {H.}~\bibnamefont
  {Tachiya}},\ }\href {https://doi.org/10.1016/S0065-2539(08)61368-2}
  {\bibfield  {journal} {\bibinfo  {journal} {Adv. Electron. Electron Phys.}\
  }\textbf {\bibinfo {volume} {28}},\ \bibinfo {pages} {337} (\bibinfo {year}
  {1969})}\BibitemShut {NoStop}%
\bibitem [{\citenamefont {Dolizy}\ \emph {et~al.}(1988)\citenamefont {Dolizy},
  \citenamefont {Groli\`ere},\ and\ \citenamefont {Lemonier}}]{Dolizy1988}%
  \BibitemOpen
  \bibfield  {author} {\bibinfo {author} {\bibfnamefont {P.}~\bibnamefont
  {Dolizy}}, \bibinfo {author} {\bibfnamefont {F.}~\bibnamefont {Groli\`ere}},\
  and\ \bibinfo {author} {\bibfnamefont {M.}~\bibnamefont {Lemonier}},\ }\href
  {https://doi.org/10.1016/S0065-2539(08)60471-0} {\bibfield  {journal}
  {\bibinfo  {journal} {Adv. Electron. Electron Phys.}\ }\textbf {\bibinfo
  {volume} {74}},\ \bibinfo {pages} {331} (\bibinfo {year} {1988})}\BibitemShut
  {NoStop}%
\bibitem [{\citenamefont {Erjavec}(1997)}]{Erjavec1997}%
  \BibitemOpen
  \bibfield  {author} {\bibinfo {author} {\bibfnamefont {B.}~\bibnamefont
  {Erjavec}},\ }\href {https://doi.org/10.1016/S0040-6090(97)00082-5}
  {\bibfield  {journal} {\bibinfo  {journal} {Thin Solid Films}\ }\textbf
  {\bibinfo {volume} {303}},\ \bibinfo {pages} {4} (\bibinfo {year}
  {1997})}\BibitemShut {NoStop}%
\bibitem [{\citenamefont {Rozhkov}\ \emph {et~al.}(2024)\citenamefont
  {Rozhkov}, \citenamefont {Bakin}, \citenamefont {Rusetsky}, \citenamefont
  {Kustov}, \citenamefont {Golyashov}, \citenamefont {Demin}, \citenamefont
  {Scheibler}, \citenamefont {Alperovich},\ and\ \citenamefont
  {Tereshchenko}}]{Rozhkov2024}%
  \BibitemOpen
  \bibfield  {author} {\bibinfo {author} {\bibfnamefont {S.~A.}\ \bibnamefont
  {Rozhkov}}, \bibinfo {author} {\bibfnamefont {V.~V.}\ \bibnamefont {Bakin}},
  \bibinfo {author} {\bibfnamefont {V.~S.}\ \bibnamefont {Rusetsky}}, \bibinfo
  {author} {\bibfnamefont {D.~A.}\ \bibnamefont {Kustov}}, \bibinfo {author}
  {\bibfnamefont {V.~A.}\ \bibnamefont {Golyashov}}, \bibinfo {author}
  {\bibfnamefont {A.~Y.}\ \bibnamefont {Demin}}, \bibinfo {author}
  {\bibfnamefont {H.~E.}\ \bibnamefont {Scheibler}}, \bibinfo {author}
  {\bibfnamefont {V.~L.}\ \bibnamefont {Alperovich}},\ and\ \bibinfo {author}
  {\bibfnamefont {O.~E.}\ \bibnamefont {Tereshchenko}},\ }\href
  {https://doi.org/10.1103/PhysRevApplied.22.024008} {\bibfield  {journal}
  {\bibinfo  {journal} {Phys. Rev. Appl.}\ }\textbf {\bibinfo {volume} {22}},\
  \bibinfo {pages} {024008} (\bibinfo {year} {2024})}\BibitemShut {NoStop}%
\bibitem [{\citenamefont {Tereshchenko}\ \emph {et~al.}(2025)\citenamefont
  {Tereshchenko}, \citenamefont {Bakin}, \citenamefont {Stepanov},
  \citenamefont {Golyashov}, \citenamefont {Mikaeva}, \citenamefont {Kustov},
  \citenamefont {Rusetsky}, \citenamefont {Rozhkov}, \citenamefont
  {Scheibler},\ and\ \citenamefont {Demin}}]{Tereshchenko2025}%
  \BibitemOpen
  \bibfield  {author} {\bibinfo {author} {\bibfnamefont {O.~E.}\ \bibnamefont
  {Tereshchenko}}, \bibinfo {author} {\bibfnamefont {V.~V.}\ \bibnamefont
  {Bakin}}, \bibinfo {author} {\bibfnamefont {S.~A.}\ \bibnamefont {Stepanov}},
  \bibinfo {author} {\bibfnamefont {V.~A.}\ \bibnamefont {Golyashov}}, \bibinfo
  {author} {\bibfnamefont {A.~S.}\ \bibnamefont {Mikaeva}}, \bibinfo {author}
  {\bibfnamefont {D.~A.}\ \bibnamefont {Kustov}}, \bibinfo {author}
  {\bibfnamefont {V.~S.}\ \bibnamefont {Rusetsky}}, \bibinfo {author}
  {\bibfnamefont {S.~A.}\ \bibnamefont {Rozhkov}}, \bibinfo {author}
  {\bibfnamefont {H.~E.}\ \bibnamefont {Scheibler}},\ and\ \bibinfo {author}
  {\bibfnamefont {A.~Y.}\ \bibnamefont {Demin}},\ }\href
  {https://doi.org/10.1103/PhysRevLett.134.157002} {\bibfield  {journal}
  {\bibinfo  {journal} {Phys. Rev. Lett.}\ }\textbf {\bibinfo {volume} {134}},\
  \bibinfo {pages} {157002} (\bibinfo {year} {2025})}\BibitemShut {NoStop}%
\bibitem [{\citenamefont {Tereshchenko}\ \emph {et~al.}(2017)\citenamefont
  {Tereshchenko}, \citenamefont {Golyashov}, \citenamefont {Rodionov},
  \citenamefont {Chistokhin}, \citenamefont {Kislykh}, \citenamefont
  {Mironov},\ and\ \citenamefont {Aksenov}}]{Tereshchenko2017}%
  \BibitemOpen
  \bibfield  {author} {\bibinfo {author} {\bibfnamefont {O.~E.}\ \bibnamefont
  {Tereshchenko}}, \bibinfo {author} {\bibfnamefont {V.~A.}\ \bibnamefont
  {Golyashov}}, \bibinfo {author} {\bibfnamefont {A.~A.}\ \bibnamefont
  {Rodionov}}, \bibinfo {author} {\bibfnamefont {I.~B.}\ \bibnamefont
  {Chistokhin}}, \bibinfo {author} {\bibfnamefont {N.~V.}\ \bibnamefont
  {Kislykh}}, \bibinfo {author} {\bibfnamefont {A.~V.}\ \bibnamefont
  {Mironov}},\ and\ \bibinfo {author} {\bibfnamefont {V.~V.}\ \bibnamefont
  {Aksenov}},\ }\href {https://doi.org/10.1038/s41598-017-16455-6} {\bibfield
  {journal} {\bibinfo  {journal} {Sci. Rep.}\ }\textbf {\bibinfo {volume}
  {7}},\ \bibinfo {pages} {16154} (\bibinfo {year} {2017})}\BibitemShut
  {NoStop}%
\bibitem [{\citenamefont {Golyashov}\ \emph {et~al.}(2020)\citenamefont
  {Golyashov}, \citenamefont {Rusetsky}, \citenamefont {Shamirzaev},
  \citenamefont {Dmitriev}, \citenamefont {Kislykh}, \citenamefont {Mironov},
  \citenamefont {Aksenov},\ and\ \citenamefont {Tereshchenko}}]{Golyashov2020}%
  \BibitemOpen
  \bibfield  {author} {\bibinfo {author} {\bibfnamefont {V.~A.}\ \bibnamefont
  {Golyashov}}, \bibinfo {author} {\bibfnamefont {V.~S.}\ \bibnamefont
  {Rusetsky}}, \bibinfo {author} {\bibfnamefont {T.~S.}\ \bibnamefont
  {Shamirzaev}}, \bibinfo {author} {\bibfnamefont {D.~V.}\ \bibnamefont
  {Dmitriev}}, \bibinfo {author} {\bibfnamefont {N.~V.}\ \bibnamefont
  {Kislykh}}, \bibinfo {author} {\bibfnamefont {A.~V.}\ \bibnamefont
  {Mironov}}, \bibinfo {author} {\bibfnamefont {V.~V.}\ \bibnamefont
  {Aksenov}},\ and\ \bibinfo {author} {\bibfnamefont {O.~E.}\ \bibnamefont
  {Tereshchenko}},\ }\href {https://doi.org/10.1016/j.ultramic.2020.113076}
  {\bibfield  {journal} {\bibinfo  {journal} {Ultramicroscopy}\ }\textbf
  {\bibinfo {volume} {218}},\ \bibinfo {pages} {113076} (\bibinfo {year}
  {2020})}\BibitemShut {NoStop}%
\bibitem [{\citenamefont {Jones}\ \emph {et~al.}(2022)\citenamefont {Jones},
  \citenamefont {Juarez-Lopez}, \citenamefont {Scheibler}, \citenamefont
  {Terekhov}, \citenamefont {Militsyn}, \citenamefont {Welsch},\ and\
  \citenamefont {Noakes}}]{Jones2022}%
  \BibitemOpen
  \bibfield  {author} {\bibinfo {author} {\bibfnamefont {L.~B.}\ \bibnamefont
  {Jones}}, \bibinfo {author} {\bibfnamefont {D.~P.}\ \bibnamefont
  {Juarez-Lopez}}, \bibinfo {author} {\bibfnamefont {H.~E.}\ \bibnamefont
  {Scheibler}}, \bibinfo {author} {\bibfnamefont {A.~S.}\ \bibnamefont
  {Terekhov}}, \bibinfo {author} {\bibfnamefont {B.~L.}\ \bibnamefont
  {Militsyn}}, \bibinfo {author} {\bibfnamefont {C.~P.}\ \bibnamefont
  {Welsch}},\ and\ \bibinfo {author} {\bibfnamefont {T.~C.~Q.}\ \bibnamefont
  {Noakes}},\ }\href {https://doi.org/10.1063/5.0109053} {\bibfield  {journal}
  {\bibinfo  {journal} {Rev. Sci. Instrum.}\ }\textbf {\bibinfo {volume}
  {93}},\ \bibinfo {pages} {113314} (\bibinfo {year} {2022})}\BibitemShut
  {NoStop}%
\bibitem [{\citenamefont {Sharma}\ \emph {et~al.}(2019)\citenamefont {Sharma},
  \citenamefont {Sreeparvathy},\ and\ \citenamefont {Kanchana}}]{Sharma2019}%
  \BibitemOpen
  \bibfield  {author} {\bibinfo {author} {\bibfnamefont {V.~K.}\ \bibnamefont
  {Sharma}}, \bibinfo {author} {\bibfnamefont {P.~C.}\ \bibnamefont
  {Sreeparvathy}},\ and\ \bibinfo {author} {\bibfnamefont {V.}~\bibnamefont
  {Kanchana}},\ }\href {https://doi.org/10.1063/1.5113282} {\bibfield
  {journal} {\bibinfo  {journal} {AIP Conf. Proc.}\ }\textbf {\bibinfo {volume}
  {2115}},\ \bibinfo {pages} {030443} (\bibinfo {year} {2019})}\BibitemShut
  {NoStop}%
\bibitem [{\citenamefont {Feng}\ \emph {et~al.}(2017)\citenamefont {Feng},
  \citenamefont {Karkare}, \citenamefont {Nasiatka}, \citenamefont {Schubert},
  \citenamefont {Smedley},\ and\ \citenamefont {Padmore}}]{Feng2017}%
  \BibitemOpen
  \bibfield  {author} {\bibinfo {author} {\bibfnamefont {J.}~\bibnamefont
  {Feng}}, \bibinfo {author} {\bibfnamefont {S.}~\bibnamefont {Karkare}},
  \bibinfo {author} {\bibfnamefont {J.}~\bibnamefont {Nasiatka}}, \bibinfo
  {author} {\bibfnamefont {S.}~\bibnamefont {Schubert}}, \bibinfo {author}
  {\bibfnamefont {J.}~\bibnamefont {Smedley}},\ and\ \bibinfo {author}
  {\bibfnamefont {H.}~\bibnamefont {Padmore}},\ }\href
  {https://doi.org/10.1063/1.4974363} {\bibfield  {journal} {\bibinfo
  {journal} {J. Appl. Phys.}\ }\textbf {\bibinfo {volume} {121}},\ \bibinfo
  {pages} {044904} (\bibinfo {year} {2017})}\BibitemShut {NoStop}%
\bibitem [{\citenamefont {Parzyck}\ \emph {et~al.}(2023)\citenamefont
  {Parzyck}, \citenamefont {Pennington}, \citenamefont {DeBenedetti},
  \citenamefont {Balajka}, \citenamefont {Echeverria}, \citenamefont {Paik},
  \citenamefont {Moreschini}, \citenamefont {Faeth}, \citenamefont {Hu},
  \citenamefont {Nangoi}, \citenamefont {Anil}, \citenamefont {Arias},
  \citenamefont {Hines}, \citenamefont {Schlom}, \citenamefont {Galdi},
  \citenamefont {Shen},\ and\ \citenamefont {Maxson}}]{Parzyck2023}%
  \BibitemOpen
  \bibfield  {author} {\bibinfo {author} {\bibfnamefont {C.~T.}\ \bibnamefont
  {Parzyck}}, \bibinfo {author} {\bibfnamefont {C.~A.}\ \bibnamefont
  {Pennington}}, \bibinfo {author} {\bibfnamefont {W.~J.~I.}\ \bibnamefont
  {DeBenedetti}}, \bibinfo {author} {\bibfnamefont {J.}~\bibnamefont
  {Balajka}}, \bibinfo {author} {\bibfnamefont {E.~M.}\ \bibnamefont
  {Echeverria}}, \bibinfo {author} {\bibfnamefont {H.}~\bibnamefont {Paik}},
  \bibinfo {author} {\bibfnamefont {L.}~\bibnamefont {Moreschini}}, \bibinfo
  {author} {\bibfnamefont {B.~D.}\ \bibnamefont {Faeth}}, \bibinfo {author}
  {\bibfnamefont {C.}~\bibnamefont {Hu}}, \bibinfo {author} {\bibfnamefont
  {J.~K.}\ \bibnamefont {Nangoi}}, \bibinfo {author} {\bibfnamefont
  {V.}~\bibnamefont {Anil}}, \bibinfo {author} {\bibfnamefont {T.~A.}\
  \bibnamefont {Arias}}, \bibinfo {author} {\bibfnamefont {M.~A.}\ \bibnamefont
  {Hines}}, \bibinfo {author} {\bibfnamefont {D.~G.}\ \bibnamefont {Schlom}},
  \bibinfo {author} {\bibfnamefont {A.}~\bibnamefont {Galdi}}, \bibinfo
  {author} {\bibfnamefont {K.~M.}\ \bibnamefont {Shen}},\ and\ \bibinfo
  {author} {\bibfnamefont {J.~M.}\ \bibnamefont {Maxson}},\ }\href
  {https://doi.org/10.1063/5.0166334} {\bibfield  {journal} {\bibinfo
  {journal} {APL Mater.}\ }\textbf {\bibinfo {volume} {11}},\ \bibinfo {pages}
  {101125} (\bibinfo {year} {2023})}\BibitemShut {NoStop}%
\bibitem [{\citenamefont {Luryi}\ and\ \citenamefont
  {Suhir}(1986)}]{Luryi1986}%
  \BibitemOpen
  \bibfield  {author} {\bibinfo {author} {\bibfnamefont {S.}~\bibnamefont
  {Luryi}}\ and\ \bibinfo {author} {\bibfnamefont {E.}~\bibnamefont {Suhir}},\
  }\href {https://doi.org/10.1063/1.97204} {\bibfield  {journal} {\bibinfo
  {journal} {Appl. Phys. Lett.}\ }\textbf {\bibinfo {volume} {49}},\ \bibinfo
  {pages} {140} (\bibinfo {year} {1986})}\BibitemShut {NoStop}%
\bibitem [{\citenamefont {Parzyck}\ \emph {et~al.}(2022)\citenamefont
  {Parzyck}, \citenamefont {Galdi}, \citenamefont {Nangoi}, \citenamefont
  {DeBenedetti}, \citenamefont {Balajka}, \citenamefont {Faeth}, \citenamefont
  {Paik}, \citenamefont {Hu}, \citenamefont {Arias}, \citenamefont {Hines},
  \citenamefont {Schlom}, \citenamefont {Shen},\ and\ \citenamefont
  {Maxson}}]{Parzyck2022}%
  \BibitemOpen
  \bibfield  {author} {\bibinfo {author} {\bibfnamefont {C.~T.}\ \bibnamefont
  {Parzyck}}, \bibinfo {author} {\bibfnamefont {A.}~\bibnamefont {Galdi}},
  \bibinfo {author} {\bibfnamefont {J.~K.}\ \bibnamefont {Nangoi}}, \bibinfo
  {author} {\bibfnamefont {W.~J.~I.}\ \bibnamefont {DeBenedetti}}, \bibinfo
  {author} {\bibfnamefont {J.}~\bibnamefont {Balajka}}, \bibinfo {author}
  {\bibfnamefont {B.~D.}\ \bibnamefont {Faeth}}, \bibinfo {author}
  {\bibfnamefont {H.}~\bibnamefont {Paik}}, \bibinfo {author} {\bibfnamefont
  {C.}~\bibnamefont {Hu}}, \bibinfo {author} {\bibfnamefont {T.~A.}\
  \bibnamefont {Arias}}, \bibinfo {author} {\bibfnamefont {M.~A.}\ \bibnamefont
  {Hines}}, \bibinfo {author} {\bibfnamefont {D.~G.}\ \bibnamefont {Schlom}},
  \bibinfo {author} {\bibfnamefont {K.~M.}\ \bibnamefont {Shen}},\ and\
  \bibinfo {author} {\bibfnamefont {J.~M.}\ \bibnamefont {Maxson}},\ }\href
  {https://doi.org/10.1103/PhysRevLett.128.114801} {\bibfield  {journal}
  {\bibinfo  {journal} {Phys. Rev. Lett.}\ }\textbf {\bibinfo {volume} {128}},\
  \bibinfo {pages} {114801} (\bibinfo {year} {2022})}\BibitemShut {NoStop}%
\end{thebibliography}
\end{document}